\documentclass[conference]{IEEEtran}

\usepackage{graphicx}
\usepackage{relsize}
\ifCLASSINFOpdf
\else
\fi

\usepackage{stfloats}
\usepackage{url}
\usepackage{xspace}
\usepackage{multicol}

\newcommand{\liquid}{Liquid\xspace}

\begin{document}
%
\title{Strong Federations: An Interoperable Blockchain Solution to Centralized Third-Party Risks}
\author{\IEEEauthorblockN{Johnny~Dilley\IEEEauthorrefmark{1},
Andrew~Poelstra\IEEEauthorrefmark{1},
Jonathan~Wilkins\IEEEauthorrefmark{1},
Marta~Piekarska\IEEEauthorrefmark{1},
Ben~Gorlick\IEEEauthorrefmark{1}, and
Mark~Friedenbach\IEEEauthorrefmark{1}}
Email: johnny, andrew, jonathan, marta, ben, mark @blockstream.com}


%


\maketitle

\begin{abstract}
Bitcoin, the first peer-to-peer electronic cash system, opened the door to permissionless, private, and trustless transactions. Attempts to repurpose Bitcoin's underlying blockchain technology have run up against fundamental limitations to privacy, faithful execution, and transaction finality. We introduce \emph{Strong Federations}: publicly verifiable, Byzantine-robust transaction networks that facilitate movement of any asset between disparate markets, without requiring third-party trust. \emph{Strong Federations} enable commercial privacy, with support for transactions where asset types and amounts are opaque, while remaining publicly verifiable. As in Bitcoin, execution fidelity is cryptographically enforced; however, \emph{Strong Federations} significantly lower capital requirements for market participants by reducing transaction latency and improving interoperability. To show how this innovative solution can be applied today, we describe \emph{\liquid}: the first implementation of \emph{Strong Federations} deployed in a Financial Market.
\end{abstract}


%

\section{Introduction}
\label{sec:intro}
\begin{figure}
\centering
\includegraphics[width=\columnwidth]{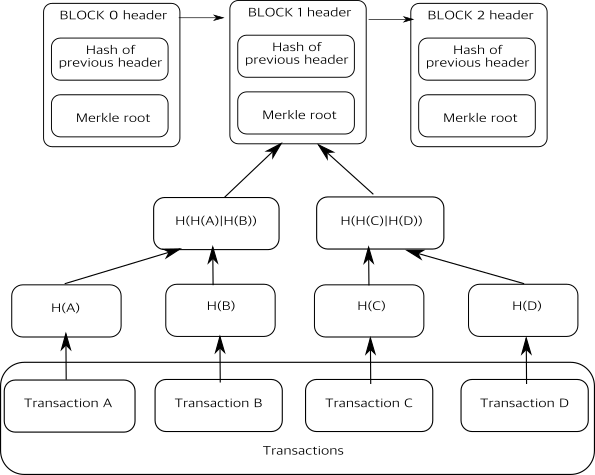}
\caption{A merkle tree connecting transactions to a block header merkle root.}
\label{fig:merkle-tree}
\end{figure}

Bitcoin, proposed by Satoshi Nakamoto in 2008, is based on the idea of a \textit{blockchain}~\cite{Nakamoto_bitcoin:a}. A blockchain consists of a series of blocks, each of which is composed of time-stamped sets of transactions and a hash of the previous block, which connects the two together, as presented in Figure~\ref{fig:merkle-tree}.

The underlying principle of Bitcoin's design is that all participants in its network are on equal footing. They jointly trust proof-of-work~\cite{Back02hashcash} to validate and enforce the network's rules, which obviates the need for central authorities such as clearinghouses. As a result, Bitcoin empowers a wide range of participants to be their own banks -- storing, transacting, and clearing for themselves without the need for a third-party intermediary. Bitcoin's network automatically enforces settlements between participants using publicly verifiable algorithms that avoid security compromises, expensive (or unavailable) legal infrastructure, third-party trust requirements, or the physical transportation of money. For the first time, users of a system have the ability to cryptographically verify other participants' behaviors, enforcing rules based on mathematics that anyone can check and no one can subvert.

Due to its design, Bitcoin has characteristics that make it a vehicle of value unlike anything that previously existed. First, it eliminates most counterparty risk from transactions~\cite{DBLP:journals/corr/NoyenVWF14}. Second, it offers cryptographic proof of ownership of assets, as the knowledge of a cryptographic key defines ownership ~\cite{poe}. Third, it is a programmable asset, offering the ability to pay to a program, or a ``smart contract", rather than a passive account or a singular public key~\cite{churchill:2015}. Fourth, and finally, it is a disruptive market mechanism for use cases such as point-to-point real-time transfers, accelerated cross-border payment, B2B remittance, asset transfers, and micropayments~\cite{Chen:2015}.

\subsection{Problem Statement}
\label{sec:problem_statement}
Because it is a global consensus system, Bitcoin's decentralized network and public verifiability come with costs. Speed of execution and insufficient guarantees of privacy are two of Bitcoin's limitations.

Bitcoin's proof-of-work methodology was designed to process transactions on average only once every ten minutes, with large variance. As a result, Bitcoin is slow from a real-time transaction processing perspective. This creates spontaneous illiquidity for parties using bitcoin\footnote{The capitalized ``Bitcoin" is used to talk about the technology and the engine, while the lowercase ``bitcoin" is used to refer to the currency.} as an intermediary, volatility exposure for those holding bitcoin for any length of time, and obstacles for the use of Bitcoin's contracting features for fast settlements. Even after a transaction is processed, counterparties must generally wait until several additional blocks have been created before considering their transaction settled. This is because Bitcoin's global ledger is at constant risk of \emph{reorganization}, wherein very recent history can be modified or rewritten. This latency undermines many commercial applications, which require real-time, or nearly instant, execution\footnote{In most traditional systems, the speed of transaction is achieved by instant execution and delayed settlement.}. Today, solving this requires a centralized counterparty, which introduces a third-party risk.

Despite issues of short-term validation, Bitcoin excels on settlement finality, providing strong assurance against transaction reversals after adequate block confirmations. In contrast, legacy payment networks leave absolute final settlement in limbo for up to 120 days typically, though chargebacks have been allowed up to 8 years late~\cite{paypal}, depending on policies imposed by the centralized network owner~\cite{FTC:CI}~\cite{FCBA}.

While a popular prevailing belief is that Bitcoin is anonymous~\cite{bloomberg}, its privacy properties are insufficient for many commercial use cases. Every transaction is published in a global ledger, which allows small amounts of information about users' financial activity (e.g., the identities of the participants in a single transaction~\cite{telegraph}) to be amplified by statistical analysis~\cite{Meiklejohn:2013:FBC:2504730.2504747}. This limits the commercial usefulness of the network and also harms individual privacy~\cite{Zohar:2015:BUH:2817191.2701411}, as user behavior frequently reflects the pervasive assumption that Bitcoin is an anonymous system. Further, it can damage the fungibility of the system, as coins that have differing histories can be identified and valued accordingly.

Overcoming these two problems would be of significance and have positive impact on the Bitcoin industry and the broader global economy~\cite{Meiklejohn:2013:FBC:2504730.2504747}. Unfortunately, previous attempts to solve similar tasks with electronic money have encountered a variety of issues: they fail to scale (e.g., BitGold~\cite{bitgold}); they are centralized (e.g., the Liberty Reserve or Wei Dai's B-money~\cite{doi:10.1080/1600910X.2013.870083}); or they raise other security concerns~\cite{DigiCash}. Moreover, higher trust requirements are often imposed through reliance on a centrally controlled system or on a single organization. This effectively replicates the problems of pre-Bitcoin systems by establishing highly permissioned arenas that have substantial regulatory disadvantages and user costs, that create onboarding and offboarding friction, and that introduce restrictions on both users and operators of the system~\cite{Nikabadi:2016:EEN:2903983.2903985}. Needless to say, if a solution is run by a central party, it is inevitably subject to systemic exposure, creating a single point of failure (SPOF) risk~\cite{BPPSH10}. The recent Ripple attack is an example of such a situation. It has been shown, that although interesting and successful otherwise, both Ripple and Stellar face the SPOF risk~\cite{rippleExploit}. Similarly, introduction of stronger trust requirements can lead to dangerous risks of consensus failure, as the consensus methods of Tendermint and Ethereum have proven~\cite{poe}. Finally, there are exchanges and brokerages that require explicit trust in a third-party~\cite{Moore2013}. Such systems leak their intrinsic insecurity into any solutions built on top of them, creating a ``house of cards" arrangement where any instability in the underlying system may result in a collapse of the dependent arrangements.

\subsection{Contributions}
This paper describes a new blockchain-based system that addresses these problems and contributes to the field in the following ways:
\begin{enumerate}
\item \textbf{Public Verifiability} -- While not fully decentralized, the system is distributed and publicly verifiable, leaving users with the ultimate spending authority over their assets.
\item \textbf{Liquidity} -- Users can move their assets into and out of the system, giving them access to its unique characteristics while also allowing them to exit at any time.
\item \textbf{No Single Point of Failure} -- The system maintains Bitcoin's permissionless innovation and avoids introducing SPOFs, all while providing novel features.
\item \textbf{Multiple Asset--Type Transfers} -- The system supports multiple asset--type transfer on the same blockchain, even within the same atomic transaction.
\item \textbf{Privacy} -- By extending earlier work on Confidential Transactions~\cite{maxwell2015} through Confidential Assets, the system supports nearly instant, trustless, atomic exchange of arbitrary goods, in a publicly verifiable but completely private way.
\item \textbf{Implementation} -- \liquid, an implementation of a Strong Federation, is presented with lessons learned from the use case of high speed inter-exchange transfers of bitcoin.
\end{enumerate}

The rest of the paper is organized as follows: the next section discusses the general design of the solution to the problems identified above -- Strong Federations. Next, in Section ~\ref{sec:technical_details} more in-depth, technical details are provided. Section~\ref{sec:use_cases} is devoted to the applications of Strong Federations in different areas. Here, \liquid is presented, the first market implementation of the system. Strong Federations are very novel in many aspects, thus some time is spent discussing various innovations in Section~\ref{sec:innovations}, to then move to a  thorough evaluation of the security and comparison of the system in Section~\ref{sec:av_tm}. Finally, Section~\ref{sec:future_work} discusses methodologies to further improve them and Section~\ref{sec:conclusions} presents conclusions.

\section{Strong Federations as a General Solution}
\label{sec:sf_gen}
\begin{figure*}
\centering
\includegraphics[width=\columnwidth]{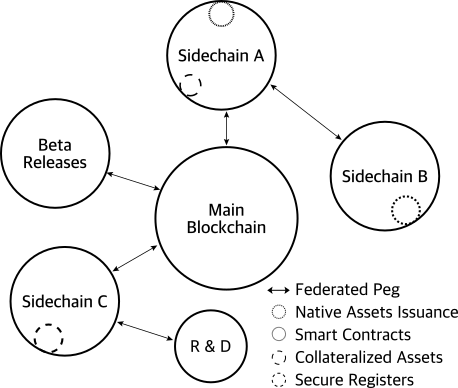}
\centering
\caption{Pegged sidechains allow parties to transfer assets by providing explicit proofs of possession in transactions.}
\label{fig:sidechains}
\end{figure*}
As mentioned in Section~\ref{sec:problem_statement}, a consensus mechanism based on proof-of-work introduces the problem of latency. However, moving to a centralized system would create significant risks of its own. To combat these problems, this paper builds on a design introduced by Back et. al. called ``Federated Pegs"~\cite{pegging}, a methodology for two-way movement of assets between Bitcoin and \emph{sidechains}. Sidechains are parallel networks that allow parties to transfer assets between blockchains by providing explicit proofs of possession within transactions, as shown in Figure~\ref{fig:sidechains}.
\subsection{Sidechains}
Sidechains are blockchains that allow users to transfer assets to and from other blockchains. At a high level, these transfers work by locking the assets in a transaction on one chain, making them unusable there, and then creating a transaction on the sidechain that describes the locked asset. Effectively, this moves assets from a parent chain to a sidechain.

This works as follows:
\begin{enumerate}
\item The user sends their asset to a special address that is designed to freeze the asset until the sidechain signals that asset is returned.
\item Using the ``in" channel of a federated peg, the user embeds information on the sidechain stating that the asset was frozen on the main chain and requests to use it on the sidechain.
\item Equivalent assets are unlocked or created on the sidechain, so that the user can participate in an alternative exchange under the sidechain rules, which can differ from the parent chain.
\item When the user wishes to move her asset, or a portion thereof, back via the ``out channel", she embeds information in the sidechain describing an output on the main blockchain.
\item The Strong Federation reaches consensus that the transaction occurred.
\item After consensus is reached, the federated peg creates such an output, unfreezing the asset on the main blockchain and assigning it as indicated on the sidechain.
\end{enumerate}

\subsection{Improving Sidechains with Strong Federations}
Bitcoin demonstrates one method of signing blocks: the use of a Dynamic Membership Multiparty Signature (DMMS) \cite{pegging} using a dynamic set of signers called \emph{miners}.
A dynamic set introduces the latency issues inherent to Bitcoin. A federated model offers another solution, with a fixed signer set, in which the DMMS is replaced with a traditional multisignature scheme.
Reducing the number of participants needed to extend the blockchain increases the speed and scalability of the system, while validation by all parties ensures integrity of the transactions. 

A \emph{Strong Federation} is a federated sidechain where the members of the federation serve as a protocol adapter between the main chain and the sidechain. One could say, essentially, that together they form a Byzantine-robust smart contract. In a Strong Federation the knowledge of private keys is sufficient for the ``right to spend" without the permission of any third-party, and the system has a mechanism that allows settlement back to a parent chain in the case of a complete failure of the federation. Not only are the code updates open and auditable and rejectable in case of coercive behavior, but the state of the system also provides a consistently reliable log that maintains immutability of state. Most importantly: the members of the federation cannot directly control any users' money inside the system other than their own.

The network operators of a Strong Federation consist of two types of \emph{functionaries}. Functionaries are entities that mechanically execute defined operations if specific conditions are met \cite{Wiki}. To enhance security, certain operations are split between entities to limit the damage an attacker can cause.  In a Strong Federation, functionaries have the power to control the transfer of assets between blockchains and to enforce the consensus rules of the sidechain. In the next section further details will be provided on why dividing those responsibilities is critical.  The two types of functionaries are:
\begin{enumerate}
\item \emph{Blocksigners}, who sign blocks of transactions on the sidechain, defining its consensus history.
\item \emph{Watchmen}, who are responsible for moving the assets out of the sidechain by signing transactions on the main chain.
\end{enumerate}
The two components can be independent.  Blocksigners are required to produce the blockchain consensus and to advance the sidechain ledger, which they do by following the protocol described in the next section. Watchmen are only required to be online when assets are to be transferred between blockchains.  As an extreme example, one could imagine a scheme where watchmen were only brought online once daily to settle a pre-approved batch of inbound and outbound transactions.

\begin{figure*}
\centering
\includegraphics[width=\columnwidth]{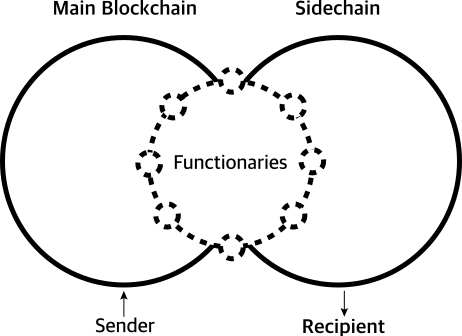}
\caption{Schematic overview of how a Strong Federation interoperates with another blockchain.}
\label{fig:strong_federations}
\end{figure*}

These two functions are performed by separate dedicated hardened boxes, configured by their owners with the secret key material required for their operation. The interaction between the elements of the network is presented in Figure~\ref{fig:strong_federations}. Between blocksigners and watchmen, only the former are required to produce  consensus, which they do by following the protocol described in the next section.

\section{Technical Details}
\label{sec:technical_details}
Supporting Strong Federations on a technical level requires the development of two types of federation: the \emph{Federated Peg} and \emph{Federated Blocksigning}.

\subsection{Federated Peg}
\label{sec:federated_peg}
The authors of ``Enabling Blockchain Innovations with Pegged Sidechains"~\cite{pegging} suggested a way to deploy federated sidechains without requiring any alterations to the consensus rules of Bitcoin's blockchain. In their methodology, a sidechain used a $k$-of-$n$ federation of mutually distrusting participants, called the functionaries, who validate and sign the blocks of the chain (blocksigners) and the pegs (watchmen) respectively.

A Federated Peg is a mechanism that uses functionaries to move assets between two chains. The functionaries observe at least the two chains -- the Bitcoin blockchain and the sidechain -- to validate asset transfers between them. To meet the criteria of a Strong Federation, a set of geographically and jurisdictionally distributed servers is used, creating a compromise-resistant network of functionaries. This network retains a number of the beneficial properties of a fully decentralized security model. 

Members of the Federated Peg each operate a secure server that runs Bitcoin and sidechain nodes along with software for creating and managing cross-chain transactions. Each server contains a hardware security module that manages cryptographic keys and signs with them. The module's job is primarily to guard the security of the network, and if a compromise is detected, to delete all of its keys, causing the network to freeze by design. If one or a few functionaries are attacked -- even if their tamper-resistant hardware is totally compromised -- the system is unaffected, as long as enough other functionaries are still intact. Successfully tampering with the federated peg system requires a compromise of at least the majority of functionaries, both blocksigners and watchmen. Even then, tampering is always detectable and usually immediately observable because the blockchain is replicated and validated on machines other than the functionaries.  A compromise of a majority of blocksigners would be observable as soon as a non-conforming block was published.  If the majority of watchmen remain secure, the value held by the sidechain can be redeemed on the parent blockchain.

\subsection{Byzantine Robustness}
\label{sec:federated_blocksigning}
One of the most important aspects of Bitcoin's mining scheme is that it is \emph{Byzantine robust}, meaning that anything short of a majority of bad actors cannot rewrite history or censor transactions~\cite{DBLP:journals/corr/ZhangDWZHW15}. The design was created to be robust against even long-term attacks of sub-majority hash rate.

Bitcoin achieves this by allowing all miners to participate on equal footing and by simply declaring that the \emph{valid} chain history with the majority of hashpower\footnote{The hashpower, or hash rate, is the measuring unit of the processing power used to secure the Bitcoin network} behind it is the true one. Would-be attackers who cannot achieve a majority are unable to rewrite history (except perhaps a few recent blocks and only with low probability) and will ultimately waste resources trying to do so. This incentivizes miners to join the honest majority, which increases the burden on other would-be attackers. However, as discussed in Section~\ref{sec:problem_statement}, this setup leads to latency due to a network heartbeat on the order of tens of minutes and introduces a risk of reorganization even when all parties are behaving honestly.

\subsection{Achieving Consensus in Strong Federations}
\label{sec:participant_selection}
It is critical that functionaries have their economic interests aligned with the correct functioning of the Federation. It would obviously be a mistake to rely on a random assortment of volunteers to support a commercial sidechain holding significant value. Beyond the incentive to attempt to extract any value contained on the sidechain, they would also have little incentive to ensure the reliability of the network. Federations are most secure when each participant has a similar amount of value held by the federation. This kind of arrangement is a common pattern in business~\cite{mohanafinancial}. Incentives can be aligned through the use of escrow, functionary allocation, or external legal constructs such as insurance policies and surety bonds.

\subsubsection{Blocksigning in Strong Federations}
In order for a Strong Federation to be low latency and eliminate the risk of reorganization from a given hostile minority, it replaces the dynamic miner set with a fixed signer set. As in Private Chains~\cite{friedenbach2013}, the validation of a script (which can change subject to fixed rules or be static) replaces the proof-of-work consensus rules. In a Strong Federation, the script implements a $k$-of-$n$ multisignature scheme. This mechanism requires blocks be signed by a certain \emph{threshold} of signers; that is, by $k$-of-$n$ signers. As such, it can emulate the Byzantine robustness of Bitcoin: a minority of compromised signers will be unable to affect the system.

\begin{figure}
\centering
\includegraphics[width=\columnwidth]{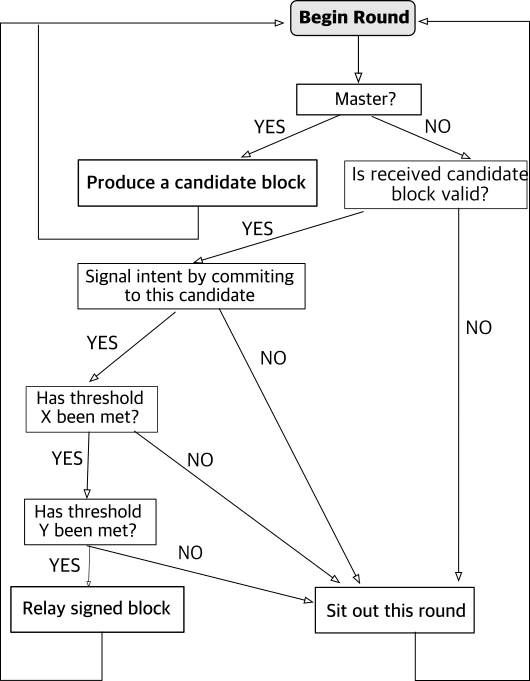}
\caption{Federated Blocksigning in a Strong Federation.}
\label{fig:blocksigner}
\end{figure}

Figure~\ref{fig:blocksigner} presents how the consensus is achieved in a  Strong Federation. It is referred to as federated blocksigning and consists of several phases:
\begin{itemize}
\item{\textbf{Step 1:}}
Blocksigners propose candidate blocks in a round-robin fashion to all other signing participants.
\item{\textbf{Step 2:}} Each blocksigner signals their intent by pre-committing to sign the given candidate block.
\item{\textbf{Step 3:}} If threshold X is met, each blocksigner signs the block.
\item{\textbf{Step 4:}} If threshold Y (which may be different from X) is met, the block is accepted and sent to the network.
\item{\textbf{Step 5:}} The next block is then proposed by the next blocksigner in the round-robin.
\end{itemize}

Due to the probabilistic generation of blocks in Bitcoin, there is a propensity for chain reorganizations in recent blocks~\cite{Eyal2014}. Because a Strong Federation's block generation is not probabilistic and is based on a fixed set of signers, it can be made to never reorganize. This allows for a significant reduction in the wait time associated with confirming transactions.

Of course, as with any blockchain-based protocol, one could imagine other ways of coordinating functionary signing. However, the proposed scheme improves the latency and liquidity of the existing Bitcoin consensus mechanism, while not introducing SPOF or higher trust requirements as discussed in Section~\ref{sec:problem_statement}.

\subsubsection{Security Improvements}
Byzantine robustness provides protection against two general classes of attack vectors. In the first case,a majority of nodes could be compromised and manipulated by the attacker, breaking integrity of the system. In the second case, a critical portion of nodes could be isolated from the network, breaking availability.

Blocksigning in a Strong Federation is robust against up to $2k - n - 1$ attackers. That is, only $2k - n$ Byzantine attackers will be able to cause conflicting blocks to be signed at the same height, forking the network. For instance, a 5-of-8 threshold would be \emph{1-Byzantine robust}\footnote{If ``Byzantine failures" in a network are caused by nodes that operate incorrectly by corrupting, forging, delaying, or sending conflicting messages to other nodes, then Byzantine robustness is defined as a network exhibiting correct behavior while a threshold of arbitrarily malfunctioning nodes (nodes with Byzantine failures) participate in the network. ~\cite{Perlman:2005:RBR:1698181}}, while 6-of-8 would be \emph{3-Byzantine robust}.

On the other hand, if at least $n - k + 1$ signers fail to sign, blocks will not be produced. Thus, increasing the threshold $k$ provides stronger protection against forks, but reduces the resilience of the network against signers being unavailable. Section~\ref{sec:upgrade_pathing} explains how the same strategy can be used for applying functionary updates, which is planned as future work.  

\section{Use Cases}
\label{sec:use_cases}
Strong Federations were developed as a technical solution to problems blockchain users face daily: transaction latency, commercial privacy, fungibility, and reliability. Many applications for blockchains require Strong Federations to avoid these issues, two of which are highlighted here. 

\subsection{International Exchange and \liquid}
Bitcoin currently facilitates remittance and cross-border payment, but its performance is hampered by technical and market dynamics~\cite{Karame:2012:DFP:2382196.2382292}. The high latency of the public Bitcoin network requires bitcoin to be tied up in multiple exchange and brokerage environments, while its limited privacy adds to the costs of operation. Due to market fragmentation, local currency trade in bitcoin can be subject to illiquidity. As a result, many commercial entities choose to operate distinct, higher-frequency methods of exchange~\cite{Moore2013}. These attempts to work around Bitcoin's inherent limitations introduce weaknesses due to centralization or other failings~\cite{Karame:2012:DFP:2382196.2382292}. 

\begin{figure*}
\centering
\includegraphics[width=0.7\textwidth]{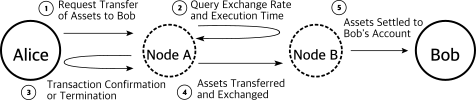}
\caption{International Trade Flow with \liquid.}
\label{fig:remittance}
\end{figure*}

We have developed a specific solution called \liquid designed to make international exchanges more efficient by utilizing bitcoin. The solution is presented in Figure~\ref{fig:remittance}, and it is the first implementation of a Strong Federation. As a Strong Federation, \liquid has novel security and trust assumptions, affording it much lower latency than Bitcoin's blockchain, with a trust model stronger than that of other, more centralized, systems (though nonetheless weaker than Bitcoin)~\cite{DBLP:journals/iacr/BanasikDM16}. Today, the implementation allows for one-minute blocks. It will be possible to reduce the time to the amount required for the pre-commit and agreement threshold time of the network traversal, as discussed in Section~\ref{sec:federated_blocksigning}. This trade-off is worthwhile in order to enable new behaviors, serving commercial needs that neither the Bitcoin blockchain nor centralized third parties can provide. 

\liquid is a Strong Federation where functionaries are exchanges participating in the network, and an asset is some currency that is transferred from Alice to Bob. As shown in Figure~\ref{fig:remittance}, when Alice wants to send money to Bob, she contacts her preferred exchange. The local node of that exchange takes care of finding an appropriate local node of the exchange willing to trade within the \liquid Strong Federation to move assets to Bob. They negotiate the terms, meaning the exchange rate and execution time, and notify Alice about the result. If she agrees, the assets are transferred to Bob. Normally, a very similar scheme would happen on the Blockchain, only the transaction would have to be approved by the whole network, thus causing horrible delay in settlement. Because \liquid operates on a sidechain, we use multisig and if 8 out of 11 participants of the Strong Federation agree on the settlement, Bob receives his money. 

A decrease in \emph{latency} in \liquid results in an increase in the speed of transaction finality. This in turn reduces the risk of bitcoin valuation changes during transaction settlement time -- a key component of successful arbitrage and remittance operations~\cite{Garcia-Alonso:2012:MMF:2169475.2169868}. The remitter will eventually receive the initial sender's bitcoin, but will have mitigated a substantial portion of the downside volatility risk by executing closer to the time of sale.

Thanks to the decrease in transfer times reducing the cost of arbitrage, \liquid participant markets will function as if they were a unified market. In addition, because \liquid assets are available at multiple fiat on- and off-ramps with relatively little delay, a remitter can settle for fiat in two or more locations in different currencies at price parity. Essentially, \liquid lowers \emph{capital constraints} relative to money held at varied end-points in the exchange cycle as a result of the network structure. 

By moving the bitcoin-holding risk, intrinsic to the operation of exchange and brokerage businesses, from a SPOF introduced by a single institution to a federation of institutions, \liquid improves the underlying security of the funds held within the network. By increasing the security of funds normally subject to explicit counterparty risk, \liquid improves the underlying reliability of the entire Bitcoin market. 

Improvements in \emph{privacy} come thanks to the adoption of Confidential Transactions, a specific addition to Strong Federations that are discussed further in Section~\ref{sec:innovations}. This provides users of the system stronger commercial privacy guarantees. 

Strong Federations such as \liquid improve privacy, latency, and reliability without exposing users to the weaknesses introduced by third-party trust. By moving business processes to \liquid, users may improve their efficiency and capital-reserve requirements.

\subsection{Other Financial Technology}
Significant portions of current financial service offerings are dependent on trusted intermediaries (and shared legal infrastructure when this trust breaks down) or centralized systems for operation~\cite{El-Yaniv:1998:CSO:274440.274442}. They have the potential to be supplanted by new, publicly verifiable consensus systems such as Bitcoin, which offer improvements to security and reliability~\cite{glantz2014:2}.

As an example, liquidity provisioning is the primary business model of Prime Brokerages and Investment Banks~\cite{gspresentation}. Fund managers commit their funds to a single location's custodianship under the premise of reducing costs associated with investment management and improving access to both investment opportunities and liquidity. Third-party broker-dealers then grant each participant access to the liquidity of their respective counterparties -- a function of aggregation of capital under a single trusted third-party custodian~\cite{wps:2013}. This system offers investors a means of preferential access to liquidity by enabling customers to buy, sell, and hedge trades with their respective counterparties in a single location. These centralized systems provide convenience to market participants, but are not without risks. One realized example of these risks is that of the Eurosystem following the global financial crisis, in the wake of the financial default of the Lehman Brothers. The effort of the Eurosystem to liquidate assets collateralized by 33 complex securities took more than four years, and resulted in over EUR 1 billion in losses~\cite{gspresentation}.

The implicit centralization and dependent trust that arise from systems like these can be resolved via a Strong Federation. It can remove the element of trust when claiming ownership and prevent transactions of uncollateralized assets, while also allowing auditing by existing and new members of the system. Furthermore, ownership of assets can be proven and verified publicly. 

\section{Innovations}
\label{sec:innovations}
In this section, major highlights of the presented design are discussed including: improvements to Determinism, Latency, and Reliability; expansions of Privacy and Confidentiality; improvements to system integrity with Hardware Security; modifications to Native Assets; and Bitcoin wallet protections via Peg-out Authorization.

\subsection{Determinism, Latency, and Reliability}
\label{sec:speed_determinism}
While Bitcoin's proof-of-work is a stochastic process, the Strong Federation scheme is deterministic, where each block is expected to be produced by a single party. Therefore \emph{reorganizations cannot happen}, unlike in Bitcoin where they are an ordinary fact of life. In a Strong Federation, blocksigners need only to obtain consensus amongst themselves before extending history; since they are a small, well-defined set, the network heartbeat can be \emph{significantly} faster than in Bitcoin. This means that users of Strong Federations can consider a single confirmation to indicate irreversibility; that confirmation can occur as quickly as information can be broadcast between the federation members and processed into a block. It also means that blocks will be produced reliably and on schedule, rather than as a stochastic process where the heartbeat is actually a mean time.

\subsection{Privacy and Confidentiality}
\label{sec:privacy}
Though many users presume that a blockchain inherently provides strong privacy, this has repeatedly been shown to be false~\cite{GervaisDecentralized}~\cite{Karame:2015:MBS:2786062.2732196}~\cite{cryptoeprint:2015:496}~\cite{DBLP:journals/iacr/SompolinskyZ13}~\cite{DBLP:conf/fc/SompolinskyZ15}. The \liquid implementation of Strong Federations uses Confidential Transactions (CT)~\cite{maxwell2015} to cryptographically verify users' behavior without providing full transparency of transaction details. As a result, the transfer of assets within a Strong Federation is guaranteed to be private between counterparties, while verifiably fair to network participants.

In order to protect confidentiality, CT blinds the amounts of all outputs to avoid leaks of information about the transaction size to third parties. It is also possible to combine inputs and outputs using traditional Bitcoin privacy techniques, such as CoinJoin~\cite{bitcointalk}. In typical application, such mechanisms are substantially weakened by the presence of public amounts~\cite{Meiklejohn:2013:FBC:2504730.2504747}, which can be used to determine mappings between owners of inputs and outputs, but in \liquid, the transaction graph no longer exposes these correlations~\cite{gibson2016}.

The use of CT in \liquid is important for two main reasons: commercial usability and fungibility. When it comes to the former, most companies would not be able to operate if their internal ledgers and financial actions were entirely public, since private business relationships and trade secrets can be inferred from transaction records. When CT is introduced this is no longer a problem as the detailed information about the trades is hidden. It is also important to improve the fungibility, because otherwise the history of an asset can be traced through the public record. This can be problematic in the case of ``tainted money", which the authorities in a given jurisdiction define as illegal or suspicious~\cite{Zdanowicz:2004:DML:986213.986239}. If an asset's history can be backtraced, then users of the network may find themselves obligated to ensure they are not receiving those assets. Such forensic work puts a large technical burden on users and operators of a network and may not even be possible across multiple jurisdictions whose definitions of taint are conflicting or ill-defined~\cite{Zdanowicz:2004:DML:986213.986239}. This is a potential danger for any type of system that enables the passing back and forth of value with a history, but one that can be corrected with improved fungibility.

Unfortunately, CT comes at a technical cost: transactions are much larger\footnote{The range of values CT can support include proofs that are often order-of-magnitude larger size than ordinary Bitcoin outputs and can be made larger depending on user requirements.} and take correspondingly longer to verify. All transactions in \liquid use CT by default, making operation of the network computationally intensive. Mimblewimble~\cite{mimblewimble} introduces a scheme by which full security may be achieved without full historical chain data, and by which transactions within blocks can no longer be distinguished from each other. This gives stronger privacy than CT alone with better scaling properties than even Bitcoin without CT. The benefits to fungibility and privacy of such a system are readily apparent. Further research will be allocated towards investigating Mimblewimble as a means of confidentially transacting. 

\subsection{Hardware Security}
\label{sec:hardware}
In Strong Federations, the $k$-of-$n$ signing requirement requires full security of the hardware, which will be distributed across multiple unknown locations and conditions. The signing keys need to be stored on the devices and not on the server for a simple reason: otherwise, even if the application code was flawless and the userspace code was minimized, a networking stack vulnerability could be exploited in order to gain access to the host and then any keys. While efforts have been made over the years to segment memory and create boundaries through virtualization, memory protection, and other means, the industry has not yet been completely successful~\cite{cve}. The best solution today is to use simplified interfaces and physical isolation; \liquid specifically creates a separate hardened device for key storage and signing in order to significantly reduce the number of avenues of attack.

While it is true that public review of cryptographic algorithms and protocols improves the security of a system, the same cannot be said for public review of hardware designs. Indeed, any measure will eventually be defeated by an attacker with an infinite supply of sample hardware. However, if a piece of hardware requires expensive, highly specialized equipment and skills to examine, it reduces the set of people who might be interested in (and capable of) attacking it. This is even more true when a technique used to break a system is destructive, requiring multiple copies of any given hardware~\cite{blackhat}.

Unfortunately, the value of hardware obfuscation for security purposes holds only until the system is broken. After an attack is published, the only way to protect the hardware is to change its design. Thus, Strong Federation hardware includes a reactive system that, when under attack, either sends an alert or simply deletes the information that it determines is likely to be targeted. Traditionally, hardware security modules do this when they register a significant environmental change such as sudden heating or cooling, a temperature out of expected operating ranges, persistent loss of access to the internet, or other environmental fluctuations~\cite{Hamdioui:2014:HPI:2616606.2616728}.

\subsection{Native Assets}
\label{sec:native_assets}
Strong Federations support accounting of other digital assets, in addition to bitcoin. These \emph{native assets} can be issued by any user and are accounted for separately from the base bitcoin currency. A participant issues such assets by means of an asset-generating transaction, optionally setting the conditions by which additional issuance can take place in the future:
\begin{enumerate}
\item The asset issuer decides on policy for the asset it is generating, including out-of-band conditions for redemption.
\item The issuer creates a transaction with one or more special \emph{asset-generating inputs}, whose value is the full issuance of the asset. This transaction, and an asset's position in it, uniquely identifies the asset. Note that the initial funds can be sent to multiple different outputs.
\item The asset-generating transaction is confirmed by a Strong Federation participant and the asset can now be transacted. The issuer distributes the asset as necessary to its customer base, using standard Strong Federation transactions.
\item Customers wishing to redeem their asset tokens transfer their asset holdings back to the issuer in return for the out-of-band good or service represented. The issuer can then destroy the tokens (i.e. by sending to an unspendable script like OP\_RETURN).
\end{enumerate}

Today, users can only trade with one asset type, however the design allows for multiple assets to be involved in a single transaction. In such cases, consensus rules ensure that the accounting equation holds true for each individual asset grouping. This allows the exchange of assets to be trustless and conducted in a single transaction without any intermediary. For that to happen, two participants who wish to trade asset A and asset B would jointly come to an agreement on an exchange rate out-of-band and produce a transaction with an A input owned by the first party and an A output owned by the second. Then the participants create another transaction with a B input owned by the second party and a B output owned by the first. This would result in a  transaction that has equal input and output amounts, and will therefore be valid if and only if both parties sign it. To finalize it, both parties would need to sign the transaction, thus executing the trade. 

What is amazing is that this innovation allows not only for exchanging currencies but any other digital assets: data, goods, information. The protocol could be further improved with more advanced sighash mechanisms.

\subsection{Peg-out Authorization}
\label{sec:whitelisting}
When moving assets from any private sidechain with a fixed membership set but stronger privacy properties than Bitcoin, it is desirable that the destination Bitcoin addresses be provably in control of some user of the sidechain. This prevents malicious or erroneous behavior on the sidechain (which can likely be resolved by the participants) from translating to theft on the wider Bitcoin network (which is irreversible).

Since moving assets back to Bitcoin is mediated by a set of watchmen, who create the transactions on the Bitcoin side, they need a dynamic private whitelist of authorized keys. That is, the members of the sidechain, who have fixed signing keys, need to be able to prove control of some Bitcoin address without associating their own identity to it, only the fact that they belong to the group. We call such proofs \emph{peg-out authorization proofs} and have accomplished it with the following design:

\begin{enumerate}
\item \textbf{Setup.} Each participant $i$ chooses two public-private keypairs: $(P_i, p_i)$ and $(Q_i, q_i)$. Here $p_i$ is an ``online key" and $q_i$ is the ``offline key". The participant gives $P_i$ and $Q_i$ to the watchmen.
\item \textbf{Authorization.} To authorize a key $W$ (which will correspond to a individually controlled Bitcoin address), a participant acts as follows.
\begin{enumerate}
\item She computes $$L_j = P_j + H(W + Q_j)(W + Q_j)$$ for every participant index $j$. Here $H$ is a random-oracle hash that maps group elements to scalars.
\item She knows the discrete logarithm of $L_i$ (since she knows the discrete logarithm of $P_i$ and chooses $W$ so she knows that of $W + Q_i$), and can therefore produce a ring signature over every $L_i$. She does so, signing the full list of online and offline keys as well as $W$.
\item She sends the resulting ring signature to the watchmen, or embeds it in the sidechain.
\end {enumerate}
\item \textbf{Transfer.} When the watchmen produce a transaction to execute transfers from the sidechain to Bitcoin, they ensure that every output of the transaction either (a) is owned by them or (b) has an authorization proof associated to its address.
\end{enumerate}

The security of this scheme can be demonstrated with an intuitive argument. First, since the authorization proofs are ring signatures over a set of keys computed identically for every participant, they are zero-knowledge for which participant produced them~\cite{Rivest2001}. Second, the equation of the keys $$L_j = P_j + H(W + Q_j)(W + Q_j)$$ is structured such that anyone signing with $L_i$ knows either:
\begin{enumerate}
\item The discrete logarithms of $W$, along with $p_i$ and $q_i$; or
\item $p_i$, but neither $q_i$ nor the discrete logarithm of $W$.
\end{enumerate}
In other words, compromise of the online key $p_i$ allows an attacker to authorize ``garbage keys" for which nobody knows the discrete logarithm. Only compromising both $p_i$ and $q_i$ (for the same $i$) will allow an attacker to authorize an arbitrary key.

However, compromise of $q_i$ is difficult because the scheme is designed such that $q_i$ need not be online to authorize $W$, as only the sum $W + Q_i$ is used when signing. Later, when $i$ wants to actually use $W$, she uses $q_i$ to compute its discrete logarithm. This can be done offline and with more expensive security requirements.

\section{Evaluation}
\label{sec:av_tm}
The information moved through Strong Federations will be very sensitive. As a result, a thorough understanding of the potential security threats is crucial. This is particularly important when dealing with Bitcoin, where transactions are irrevocable. In other words, continued operation of the network is a secondary priority; few would choose to have their money move rapidly into the hands of a thief over a delayed return to their own pockets. As the aggregate value of assets inside a Strong Federation increases, the incentives for attackers grow, and it becomes crucial they cannot succeed when targeting any functionary nor the maintainer of the codebase. Thankfully, as participants in a Strong Federation scale up the value of assets flowing through the system, they will be naturally incentivized to take greater care of access to the federated signers under their control. Thus, the federated security model neatly aligns with the interests of its participants.

\subsection{Comparison to Existing Solutions}
Existing proposals to form consensus for Bitcoin-like systems generally fall into two categories: those that attempt to preserve Bitcoin's decentralization while improving efficiency or throughput and those that adopt a different trust model altogether. In the first category are GHOST~\cite{DBLP:journals/corr/SompolinskyZ16}, block DAGs~\cite{DBLP:conf/atal/LewenbergBSZR15}~\cite{DBLP:journals/corr/Kokoris-KogiasJ16}, and Jute~\cite{jute}. These schemes retain Bitcoin's model of blocks produced by a dynamic set of anonymous miners, and depend on complex and subtle game-theoretic assumptions to ensure consensus is maintained in a decentralized way. The second category includes schemes such as Stellar~\cite{stellar} and Tendermint~\cite{tendermint}, which require new participants to choose existing ones to trust. These examples have the failure risks associated with trusted parties, which when spread across complex network topologies lead to serious but difficult-to-analyze failure modes~\cite{ycom}.

Our proposal works in the context of a fixed set of mutually distrusting but identifiable parties, and therefore supports a simple trust model: as long as a quorum of participants act honestly, the system continues to work.

Parallel to consensus systems are systems that seek to leverage existing consensus systems to obtain faster and cheaper transaction execution. The primary example of this is the Lightning network~\cite{lightning}, which allows parties to transact by interacting solely with each other, only falling back to the underlying blockchain during a setup phase or when one party fails to follow the protocol. We observe that since these systems work on top of existing blockchains, they complement new consensus systems, including the one described in this paper.

A novel proposal has been presented recently by Eyal et al.~\cite{194906}. Although not yet available on the market, their Bitcoin-NG scheme is a new blockchain protocol designed to scale. Based on the experiments they conduct it seems like their solution scales optimally, with bandwidth limited only by the capacity of the individual nodes and latency limited only by the propagation time of the network. However, there may be game-theoretic failings or denial-of-service vectors inherent to their design that have yet to be explicated.

\subsection{Protection Mechanisms}
\label{sec:protections}
An attacker must first communicate with a system to attack it, so the communication policies for a Strong Federation have been designed to isolate it from common attack vectors. Several different measures are taken to prevent untrustworthy parties from communicating with functionaries:
\begin{itemize}
\item Functionary communications are restricted to hard-coded Tor Hidden Service addresses known to correspond to known-peer functionaries.
\item Inter-functionary traffic is authenticated using hard-coded public keys and per-functionary signing keys.
\item The use of Remote Procedure Calls (RPC) is restricted on functionary hardware and on \liquid wallet deployments to callers on the local system only.
\end{itemize}

Above and beyond, the key policy works to protect the network. While the blocksigners are designed with secret keys that are unrecoverable in any situation, the watchmen keys must be created with key recovery processes in mind. Loss of the blocksigner key would require a hard-fork of the Strong Federation's consensus protocol. This, while difficult,  is possible and does not risk loss of funds. However, loss of sufficient watchmen keys would result in the loss of bitcoin and is unacceptable.

Although the Strong Federation design is Byzantine robust, it is still very important that functionaries avoid compromise. Given tamper-evident sensors designed to detect attacks on functionaries, if an attack is determined to be in progress, it is important to inform other functionaries in the network that its integrity can no longer be guaranteed. In this case, the fallback is to shutdown the individual system, and in a worst case scenario, where the Byzantine robustness of the network is potentially jeopardized, the network itself should shutdown. This ensures a large safety margin against system degradation -- assuring both the direct security of users' funds and the users' confidence in the system's continued correct operation.

\subsection{Backup Withdrawal}
\label{sec:unilateral_withdrawal}
The blockchain for the \liquid implementation of Strong Federations is publicly verifiable, and it should be possible, in principle, for holders of bitcoin in \liquid to move their coins back to Bitcoin even under conditions where the \liquid network has stalled (due to DoS or otherwise).

The most straightforward way of doing this would be for watchmen to provide time-locked Bitcoin transactions, returning the coins to their original owners. However, this updates the recipient-in-case-of-all-stall only at the rate that time-locked transactions are invalidated, which may be on the order of hours or days. The actual owners of coins on the \liquid chain will change many times in this interval, so this solution does not work. Bitcoin does not provide a way to prove ownership with higher resolution than this.

However, it is possible to set a ``backup withdrawal address" that is controlled by a majority of network participants, functionaries, and external auditors. This way, if \liquid stalls, it is possible for affected parties to collectively decide on appropriate action.

\subsection{Availability and Denial of Service}
\label{sec:availability}
There are two independent thresholds involved when signing blocks in a Strong Federation: the signing and precommit thresholds. The former is an unchangeable property of the network and may be set with resilience in mind. It may also be adjusted to a more advanced policy that supports backup blocksigners that are not normally online. The precommit threshold, on the other hand, is determined only by the signers themselves and may be set to a high level (even requiring unanimity of signers) and changed as network conditions require. This means that even if the network block signature rules in principle allow Byzantine attackers to cause forks, in practice malicious users are (at worst) limited to causing a denial of service to the network, provided that the blocksigners set a high enough precommit threshold.

A software bug or hardware failure could lead to a breakdown in a single functionary such that it temporarily no longer functions. Such a participant would no longer be able to take part in the consensus protocol or be able to approve withdrawals to Bitcoin. Unless enough functionaries fail so that the signing threshold is not achievable, the network will be unaffected\footnote{Downtime for a functionary does cause degradation in throughput performance as that signer's turn will have to be ``missed" each round.}. In such a case, funds will be unable to move (either within the sidechain or back to Bitcoin) for the duration of the outage. Once the functionaries are restored to full operation, the network will continue operating, with no risk to funds.

\subsection{Hardware Failure}
\label{sec:hardware_failure}
If a blocksigner suffers hardware failure, and cryptographic keys are not recoverable, the entire network must agree to change signing rules to allow for a replacement blocksigner. 

A much more serious scenario involves the failure of a watchman, as its keys are used by the Bitcoin network and cannot be hard-forked out of the current Bitcoin signature set. If a single watchman fails, it can be replaced and the other watchmen will be able to move locked coins to ones protected by the new watchman's keys. However, if too many watchmen fail at once, and their keys are lost, bitcoin could become irretrievable. As mentioned in Section \ref{sec:unilateral_withdrawal}, this risk can be mitigated by means of a backup withdrawal mechanism.

Prevention mechanisms include extraction and backup withdrawal of watchman key material, so that bitcoin can be recovered in the event of such a failure. The encryption of extracted keys ensures that they can only be seen by the original owner or an independent auditor. This prevents individual watchmen operators (or anyone with physical access to the watchmen) from extracting key material that could be used to operate outside of the sidechain.

\subsection{Rewriting History}
\label{sec:forking}
It is possible that blocksigners could attempt to rewrite history by forking a Strong Federation blockchain. Compared to Bitcoin, it is quite cheap to sign conflicting histories if one is in possession of a signing key.

However, rewriting the chain would require compromising the keys held in secure storage on a majority of blocksigners. Such an attack is an unlikely scenario, as it would require determining the locations of several signers, which are spread across multiple countries in multiple continents, and either bypassing the tamper-resistant devices or else logically accessing keys through an exploit of the underlying software.

Further, such an attack is detectable, and a proof (consisting simply of the headers of the conflicting blocks) can be published by anybody and used to automatically stop network operation until the problem is fixed and compromised signers replaced.

If the network was forked in this way, it might be possible for active attackers to reverse their own spending transactions by submitting conflicting transactions to both sides of the fork. Therefore, any “valid” blocks that are not unique in height should be considered invalid.

\subsection{Transaction Censorship}
\label{sec:censorship}
By compromising a threshold number of blocksigners, an attacker can potentially enforce selective transaction signing by not agreeing to sign any blocks that have offending transactions and not including them in their own proposed blocks. Such situations might occur due to a conflict between legitimate signers or the application of legal or physical force against them.

This type of censorship is not machine-detectable, although it may become apparent that specific blocksigners are being censored if they have many unsuccessful proposal rounds. It will be obvious to the affected network participants that something is happening, and in this case the Strong Federation may use the same mechanism to replace or remove the attacking signers that is used to resolve other attacks.

\subsection{Confiscation of Locked Bitcoins}
\label{sec:confiscation}
If enough watchmen collude, they can overcome the multisig threshold and confiscate all the bitcoin currently in the sidechain.

The resilience against such attacks can be improved by setting a high signing threshold on the locked bitcoins. This can exclude all but the most extreme collusion scenarios. However, this weakens resilience against failures of the watchmen whose key material is lost. The cost-benefit analysis will have to be done as federated signing technology matures.

\section{Future Research}
\label{sec:future_work}
While Strong Federations introduce new technology to solve a variety of longstanding problems, these innovations are far from the end of the road. The ultimate design goal is to have a widely distributed network in which the operators are physically unable to interfere or interact with application-layer processes in any way, except possibly by entirely ceasing operation, with backup plans to retrieve the funds to the parent chain.

\subsection{Further Hardening of Functionaries}
\label{sec:further_hardening}
More research should be done to ensure that functionaries cannot be physically tampered with, and that network interactions are legitimate and auditable. Methods for future hardening could include specific design improvements or further cryptographic arrangements. In a Strong Federation, compromised functionaries are unable to steal funds, reverse transactions, or influence other users of the system in any way. However, enough malicious functionaries can always stall the network by refusing to cooperate with other functionaries or by shutting down completely. This could freeze funds until an automatic withdrawal mechanism starts.

As such, it would be beneficial to research possibilities for creating incentive structures and methods to encourage functionary nodes to remain online under attack. This could be done, for example, by requiring that they periodically sign time-locked transactions. These incentives could prevent certain denial-of-service attacks.

\subsection{Enlightening Liquid}
\label{sec:enlightening_liquid}
The privacy and speed of a Strong Federation could be further improved by combining it with Lightning~\cite{lightning}. Just as with Bitcoin's network, the throughput of initial systems built with this architecture is limited on purpose, as the transactions are published in blocks that must be made visible to all participants in the network. This threshold is set by the need for everyone to see and validate each operation. Even with a private network that mandates powerful hardware, this is a serious drawback.

With Lightning, individual transactions only need to be validated by the participating parties~\cite{lightning}. This dramatically reduces the verification load for all participants. Because end-to-end network speed is the only limiting factor~\cite{DBLP:conf/sss/DeckerW15}, it also greatly reduces the effects of network latency.

Furthermore, nodes in a Strong Federation could route payments via Lightning, a network of bidirectional payment channel smart contracts. This may allow for even more efficient entry and exit from the \liquid network. Finally, Lightning can replace inter-chain atomic swap smart contracts and probably hybrid multi-chain transitive trades~\cite{friedenbach2013} without having the limitation of a single DMMS chain.

\subsection{Confidential Assets}
\label{sec:confidential_assets}
Confidential Transactions (CT) hide the \emph{amounts} but not the \emph{types} of transacted assets, so its privacy is not as strong as it could be. However, CT could be extended to also hide the asset type. For any transaction it would be impossible to determine which assets were transacted or in what amounts, except by the parties to the transaction. Called \emph{Confidential Assets} (CA), this technology improves user privacy and allows transactions of unrelated asset classes to be privately spent in a single transaction. However, the privacy given to assets is qualitatively different than that of CT.

Consider a transaction with inputs of asset types A and B. All observers know that the outputs have types A and B, but they are unable to determine which outputs have what types (or how they are split up or indeed anything about their amounts). Therefore all outputs of this transaction will have type ``maybe A, maybe B" from the perspective of an observer. Suppose then that an output of type A, which is a ``maybe A, maybe B" to those not party to the transaction, is later spent in a transaction with an asset type C. The outputs of this transaction would then be ``maybe A, maybe B, maybe C" to outside observers, ``maybe A, maybe C" to those party to the first transaction, and known to those party to the second.

As transactions occur, outputs become increasingly ambiguous as to their type, except to individual transactors, who know the true types of the outputs they own. If issuing transactions always have multiple asset--types, then non-participant observers never learn the true types of any outputs.
\subsection{Byzantine-Robust Upgrade Paths}
\label{sec:upgrade_pathing}
Most hardening approaches rely on a central, trusted third-party who can provide upgrades: operating systems and other critical software wait for signed software packages, generated in locked down build labs, then hosts retrieve these packages, verify signatures, and apply them, often automatically. This would undermine the threat model of a Strong Federation, as any SPOF can be compromised or coerced to comply. All aspects of the change control system must instead be defensibly Byzantine secure. In any large system, one must assume some part of it may be in a state of failure or attack at any point in time. This means that what can be a simple process for a central authority becomes somewhat more complex.

Unfortunately, creating an agile network, or a system that is upgradable, requires a security tradeoff. An ideal balance is hard to strike: as a network's independence grows, the cost and difficulty in upgrading also increases. As such, it is important that all changes in the code should be opt-in for all parties and the process should be consensus ($k$-of-$n$) driven across the functionary set. These changes should also be fully auditable and transparent prior to application.

Ultimately, the processes of maintenance, of new member additions, or of strict improvements to the network must also be Byzantine secure for the whole system to be Byzantine secure. For Bitcoin, this is achieved by a long-tail upstream path which is an audited and open-source procedure, and ultimately the consensus rules each user decides to validate are self-determined (i.e., there may be permanent chain splits in case of controversial changes).
\begin{figure*}[ht]
\centering
\includegraphics[width=0.7\textwidth]{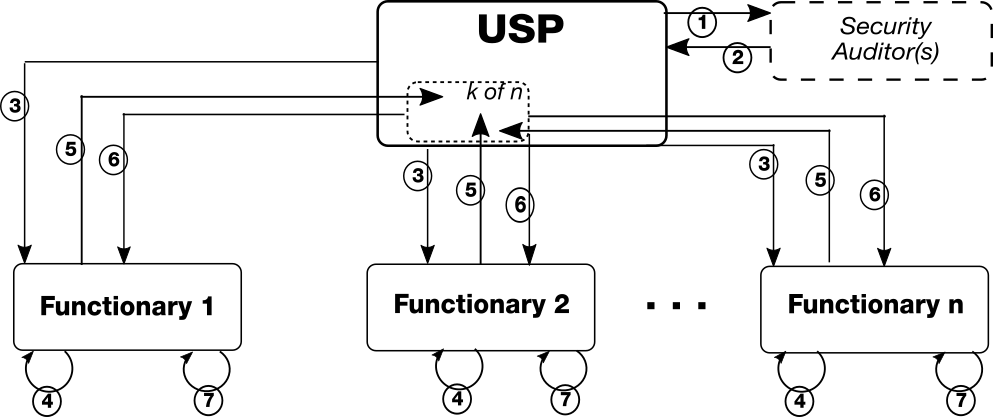}
\centering\caption{Byzantine-Robust Upgrade schema planned for Strong Federations.}
\label{fig:br}
\end{figure*}

For Strong Federations, this will be achieved through the design and implementation of an upgrade procedure that enables iterative improvement to the system without enabling attack surfaces by emulating Bitcoin's soft-fork upgrade path. This is presented in Figure~\ref{fig:br}, and follows the steps:

\begin{enumerate}
\item An upstream software provider (USP) writes software updates for the functionary network and provides those updates to the functionaries for implementation.
\item An external security auditor may be used to review the software update and documentation for correctness, verifying the accuracy of the documentation and/or the codebase itself.
\item Each functionary verifies the signatures from the USP and possibly the third-party auditor, and may also review or audit the updates if it wishes.
\item Each functionary signs the update on the server and returns the resulting signature to the USP.
\item Once a supermajority of functionaries have signed the update, the USP combines their signatures and the update image into a single package. This file, consisting of the update image, documentation, and a supermajority of functionary signatures, is then distributed to each functionary.
\item Each functionary receives the USP and supermajority signatures on their server. 
\item Each functionary verifies the package contents and applies the update.
\end{enumerate}

Note that this situation assumes honest participants. There are scenarios in which, for instance, a single group of collaborating malicious functionaries can collectively reject any given upgrade path. Methods of combating this scenario will be further investigated.

\section{Conclusion}
\label{sec:conclusions}
The popularity of Bitcoin shows that permissionless proof-of-work is an effective mechanism for developing an infrastructure, with hundreds of millions of dollars~\cite{Sirer:2016:TPS:2907055.2896382} across dozens of companies being invested in new innovations spanning chip and network design, datacenter management, and cooling systems~\cite{Andrychowicz:2016:SMC:2907055.2896386}. The value of the security offered by this conglomerate of resources is immense. There is, however, a drawback of the proof-of-work underlying Bitcoin~\cite{Back02hashcash}: the addition of latency (the block time) to establish widely distributed checkpoints for the shared, current state of the ledger.

This paper introduces the Strong Federation: a federated consensus mechanism which significantly mitigates a number of real-world systemic risks when used in conjunction with proof-of-work. The solution is resilient against broad categories of attacks via specific implementation decisions and minimization of attack surfaces. Strong Federations improve blockchain technology by leveraging sidechain technology. Furthermore, market enhancements utilizing Confidential Transactions and Native Assets are proposed.

This paper proposes a methodology that utilizes hardware security modules (HSMs)  for enforcing consensus. Currently HSMs have limited ability to verify that their block signatures are only used on valid histories that do not conflict with past signatures. This arises both because of the performance limitations in secure hardware and because anything built into an HSM becomes unchangeable, making complex rule sets difficult \emph{and risky} to deploy. Improved verification requires HSMs to support an upgrade path that is sufficiently capable while being hardened against non-authorized attempts at upgrading. Alternatively, every software deployment may imply a new hardware HSM deployment, but that's not cost efficient.

The first working implementation of a Strong Federation is \liquid~--~a Bitcoin exchange and brokerage multi-signature sidechain that bypasses Bitcoin's inherent limitations while leveraging its security properties. In \liquid, Bitcoin's proof-of-work is replaced with a $k$-of-$n$ multisignature scheme. In this model, consensus history is a blockchain where every block is signed by the majority of a deterministic, globally distributed set of functionaries running on hardened platforms, a methodology that directly aligns incentives for the participants.

Strong Federations will be useful in many general-purpose industries -- especially those that seek to represent and exchange their assets digitally and must do so securely and privately without a single party that controls the custodianship, execution, and settling of transactions.

\section*{Acknowledgements}
We thank Matt Corallo and Patrick Strateman for their substantial commentary and contribution in the formation of the ideas and process behind this paper. We'd also like to thank Eric Martindale, Jonas Nick, Greg Sanders, and Kat Walsh for their extensive review. Finally, we thank Kiara Robles for her excellent figures and diagrams.



\begin{thebibliography}{10}
\providecommand{\url}[1]{#1}
\csname url@samestyle\endcsname
\providecommand{\newblock}{\relax}
\providecommand{\bibinfo}[2]{#2}
\providecommand{\BIBentrySTDinterwordspacing}{\spaceskip=0pt\relax}
\providecommand{\BIBentryALTinterwordstretchfactor}{4}
\providecommand{\BIBentryALTinterwordspacing}{\spaceskip=\fontdimen2\font plus
\BIBentryALTinterwordstretchfactor\fontdimen3\font minus
  \fontdimen4\font\relax}
\providecommand{\BIBforeignlanguage}[2]{{%
\expandafter\ifx\csname l@#1\endcsname\relax
\typeout{** WARNING: IEEEtran.bst: No hyphenation pattern has been}%
\typeout{** loaded for the language `#1'. Using the pattern for}%
\typeout{** the default language instead.}%
\else
\language=\csname l@#1\endcsname
\fi
#2}}
\providecommand{\BIBdecl}{\relax}
\BIBdecl

\bibitem{Nakamoto_bitcoin:a}
\BIBentryALTinterwordspacing
S.~Nakamoto, ``Bitcoin: A peer-to-peer electronic cash system,'' 2009.
  [Online]. Available: \url{http://bitcoin.org/bitcoin.pdf}
\BIBentrySTDinterwordspacing

\bibitem{Back02hashcash}
A.~Back, ``Hashcash - a denial of service counter-measure,'' Tech. Rep., 2002.

\bibitem{DBLP:journals/corr/NoyenVWF14}
\BIBentryALTinterwordspacing
K.~Noyen, D.~Volland, D.~W{\"{o}}rner, and E.~Fleisch, ``When money learns to
  fly: Towards sensing as a service applications using bitcoin,'' \emph{CoRR},
  vol. abs/1409.5841, 2014. [Online]. Available:
  \url{http://arxiv.org/abs/1409.5841}
\BIBentrySTDinterwordspacing

\bibitem{poe}
\BIBentryALTinterwordspacing
B.~Wiki, ``Proof of ownership.'' [Online]. Available:
  \url{https://en.bitcoin.it/wiki/Proof_of_Ownership}
\BIBentrySTDinterwordspacing

\bibitem{churchill:2015}
E.~F. Churchill., ``Why should we care about bitcoin?'' \emph{Interactions},
  2015.

\bibitem{Chen:2015}
W.~K.~H. Dengke~Chen, Shiyi~Chen, ``Internet finance: Digital currencies and
  alternative finance liberating the capital markets,'' \emph{JOURNAL OF
  GOVERNANCE AND REGULATION}, vol. 4, Issue 4, pp. 190--202, 2015.

\bibitem{paypal}
\BIBentryALTinterwordspacing
 [Online]. Available:
  \url{http://webcache.googleusercontent.com/search?q=cache:-coDmwI95UYJ:community.ebay.com/t5/Archive-Trust-Safety-Safe-Harbor/Paypal-Chargeback-8-YEARS-after-sale/td-p/16473845+&cd=15&hl=en&ct=clnk&gl=us}
\BIBentrySTDinterwordspacing

\bibitem{FTC:CI}
{Federal Trade Comission}, ``Disputing credit card charges,'' \emph{Consumer
  Information}, 2016.

\bibitem{FCBA}
\BIBentryALTinterwordspacing
------, ``Fair credit billing act,'' in \emph{Public Law 93-495}.\hskip 1em
  plus 0.5em minus 0.4em\relax 93rd Congress - H.R. 11221, 15 USC 1601 July 9,
  1986. [Online]. Available:
  \url{https://www.ftc.gov/sites/default/files/fcb.pdf}
\BIBentrySTDinterwordspacing

\bibitem{bloomberg}
\BIBentryALTinterwordspacing
 [Online]. Available:
  \url{http://www.bloomberg.com/news/articles/2013-10-01/lawsky-says-so-be-it-if-transparency-dooms-bitcoin}
\BIBentrySTDinterwordspacing

\bibitem{telegraph}
\BIBentryALTinterwordspacing
 [Online]. Available:
  \url{http://www.telegraph.co.uk/technology/news/11451379/US-auctions-50000-Bitcoins-seized-from-Silk-Road.html}
\BIBentrySTDinterwordspacing

\bibitem{Meiklejohn:2013:FBC:2504730.2504747}
\BIBentryALTinterwordspacing
S.~Meiklejohn, M.~Pomarole, G.~Jordan, K.~Levchenko, D.~McCoy, G.~M. Voelker,
  and S.~Savage, ``A fistful of bitcoins: Characterizing payments among men
  with no names,'' in \emph{Proceedings of the 2013 Conference on Internet
  Measurement Conference}, ser. IMC '13.\hskip 1em plus 0.5em minus 0.4em\relax
  New York, NY, USA: ACM, 2013, pp. 127--140. [Online]. Available:
  \url{http://doi.acm.org/10.1145/2504730.2504747}
\BIBentrySTDinterwordspacing

\bibitem{Zohar:2015:BUH:2817191.2701411}
\BIBentryALTinterwordspacing
A.~Zohar, ``Bitcoin: Under the hood,'' \emph{Commun. ACM}, vol.~58, no.~9, pp.
  104--113, Aug. 2015. [Online]. Available:
  \url{http://doi.acm.org/10.1145/2701411}
\BIBentrySTDinterwordspacing

\bibitem{bitgold}
T.~Mayer, ``Bitgold review,'' \emph{The Great Credit Contraction}, 2014.

\bibitem{doi:10.1080/1600910X.2013.870083}
\BIBentryALTinterwordspacing
H.~Karlstr{\o}m, ``Do libertarians dream of electric coins? the material
  embeddedness of bitcoin,'' \emph{Distinktion: Journal of Social Theory},
  vol.~15, no.~1, pp. 23--36, 2014. [Online]. Available:
  \url{http://dx.doi.org/10.1080/1600910X.2013.870083}
\BIBentrySTDinterwordspacing

\bibitem{DigiCash}
\BIBentryALTinterwordspacing
Unknown, ``How digicash blew everything,'' \emph{Next! Magazine}, 1999.
  [Online]. Available: \url{http://www.nextmagazine.nl/ecash.htm}
\BIBentrySTDinterwordspacing

\bibitem{Nikabadi:2016:EEN:2903983.2903985}
\BIBentryALTinterwordspacing
M.~S. Nikabadi and S.~M. Mousavi, ``The effect of e-money on the non-financial
  performance of banks: Case study: Bank mellat of iran,'' \emph{Int. J. Innov.
  Digit. Econ.}, vol.~7, no.~1, pp. 12--23, Jan. 2016. [Online]. Available:
  \url{http://dx.doi.org/10.4018/IJIDE.2016010102}
\BIBentrySTDinterwordspacing

\bibitem{BPPSH10}
\BIBentryALTinterwordspacing
S.~V. Buldyrev, R.~Parshani, G.~Paul, H.~E. Stanley, and S.~Havlin,
  ``Catastrophic cascade of failures in interdependent networks,''
  \emph{Nature}, vol. 464, no. 7291, pp. 1025--1028, Apr. 2010. [Online].
  Available: \url{http://dx.doi.org/10.1038/nature08932}
\BIBentrySTDinterwordspacing

\bibitem{rippleExploit}
D.~Hide, ``Exploiting ripple transaction ordering for fun and profit,''
  Available Imagination, Tech. Rep., 2015.

\bibitem{Moore2013}
\BIBentryALTinterwordspacing
T.~Moore and N.~Christin, \emph{Beware the Middleman: Empirical Analysis of
  Bitcoin-Exchange Risk}.\hskip 1em plus 0.5em minus 0.4em\relax Berlin,
  Heidelberg: Springer Berlin Heidelberg, 2013, pp. 25--33. [Online].
  Available: \url{http://dx.doi.org/10.1007/978-3-642-39884-1_3}
\BIBentrySTDinterwordspacing

\bibitem{maxwell2015}
\BIBentryALTinterwordspacing
G.~Maxwell, 2015. [Online]. Available:
  \url{https://people.xiph.org/~greg/confidential_values.txt}
\BIBentrySTDinterwordspacing

\bibitem{pegging}
A.~Back, M.~Corallo, L.~Dashjr, M.~Friedenbach, G.~Maxwell, A.~Miller,
  A.~Poelstra, J.~Tim{\'o}n, and P.~Wuille, ``Enabling blockchain innovations
  with pegged sidechains,'' 2014.

\bibitem{Wiki}
\BIBentryALTinterwordspacing
 [Online]. Available: \url{https://en.wiktionary.org/wiki/functionary}
\BIBentrySTDinterwordspacing

\bibitem{DBLP:journals/corr/ZhangDWZHW15}
\BIBentryALTinterwordspacing
L.~Zhang, G.~Ding, Q.~Wu, Y.~Zou, Z.~Han, and J.~Wang, ``Byzantine attack and
  defense in cognitive radio networks: {A} survey,'' \emph{CoRR}, vol.
  abs/1504.01185, 2015. [Online]. Available:
  \url{http://arxiv.org/abs/1504.01185}
\BIBentrySTDinterwordspacing

\bibitem{mohanafinancial}
\BIBentryALTinterwordspacing
R.~Mohana, \emph{Financial Statement Analysis and Reporting}.\hskip 1em plus
  0.5em minus 0.4em\relax PHI Learning Pvt. Ltd. [Online]. Available:
  \url{https://books.google.com/books?id=97a8SG9hvMgC}
\BIBentrySTDinterwordspacing

\bibitem{friedenbach2013}
\BIBentryALTinterwordspacing
M.~Friedenbach and J.~Tim\'on, ``Freimarkets: extending bitcoin protocol with
  user-specified bearer instruments, peer-to-peer exchange, off-chain
  accounting, auctions, derivatives and transitive transactions.'' [Online].
  Available: \url{http://freico.in/docs/freimarkets-v0.0.1.pdf}
\BIBentrySTDinterwordspacing

\bibitem{Eyal2014}
\BIBentryALTinterwordspacing
I.~Eyal and E.~G. Sirer, \emph{Financial Cryptography and Data Security: 18th
  International Conference, FC 2014, Christ Church, Barbados, March 3-7, 2014,
  Revised Selected Papers}.\hskip 1em plus 0.5em minus 0.4em\relax Berlin,
  Heidelberg: Springer Berlin Heidelberg, 2014, ch. Majority Is Not Enough:
  Bitcoin Mining Is Vulnerable, pp. 436--454. [Online]. Available:
  \url{http://dx.doi.org/10.1007/978-3-662-45472-5_28}
\BIBentrySTDinterwordspacing

\bibitem{Perlman:2005:RBR:1698181}
R.~Perlman, ``Routing with byzantine robustness,'' Mountain View, CA, USA,
  Tech. Rep., 2005.

\bibitem{Karame:2012:DFP:2382196.2382292}
\BIBentryALTinterwordspacing
G.~O. Karame, E.~Androulaki, and S.~Capkun, ``Double-spending fast payments in
  bitcoin,'' in \emph{Proceedings of the 2012 ACM Conference on Computer and
  Communications Security}, ser. CCS '12.\hskip 1em plus 0.5em minus
  0.4em\relax New York, NY, USA: ACM, 2012, pp. 906--917. [Online]. Available:
  \url{http://doi.acm.org/10.1145/2382196.2382292}
\BIBentrySTDinterwordspacing

\bibitem{DBLP:journals/iacr/BanasikDM16}
\BIBentryALTinterwordspacing
W.~Banasik, S.~Dziembowski, and D.~Malinowski, ``Efficient zero-knowledge
  contingent payments in cryptocurrencies without scripts,'' \emph{{IACR}
  Cryptology ePrint Archive}, vol. 2016, p. 451, 2016. [Online]. Available:
  \url{http://eprint.iacr.org/2016/451}
\BIBentrySTDinterwordspacing

\bibitem{Garcia-Alonso:2012:MMF:2169475.2169868}
\BIBentryALTinterwordspacing
C.~R. Garc\'{\i}a-Alonso, E.~Arenas-Arroyo, and G.~M. P{\'e}rez-Alcal\'{a}, ``A
  macro-economic model to forecast remittances based on monte-carlo simulation
  and artificial intelligence,'' \emph{Expert Syst. Appl.}, vol.~39, no.~9, pp.
  7929--7937, Jul. 2012. [Online]. Available:
  \url{http://dx.doi.org/10.1016/j.eswa.2012.01.108}
\BIBentrySTDinterwordspacing

\bibitem{El-Yaniv:1998:CSO:274440.274442}
\BIBentryALTinterwordspacing
R.~El-Yaniv, ``Competitive solutions for online financial problems,'' \emph{ACM
  Comput. Surv.}, vol.~30, no.~1, pp. 28--69, Mar. 1998. [Online]. Available:
  \url{http://doi.acm.org/10.1145/274440.274442}
\BIBentrySTDinterwordspacing

\bibitem{glantz2014:2}
\BIBentryALTinterwordspacing
R.~Glantz and J.~Dilley, ``The bitcoin blockchain,'' Tech. Rep., 2014.
  [Online]. Available:
  \url{https://cdn.panteracapital.com/wp-content/uploads/Bitcoin-Blockchain-White-Paper1.pdf}
\BIBentrySTDinterwordspacing

\bibitem{gspresentation}
L.~C. Blankfein, ``Goldman sachs presentation to credit suisse financial
  services conference,'' Goldman Sachs, Tech. Rep., 2015.

\bibitem{wps:2013}
U.~Bindseil and J.~Jab{\l}ecki, ``Working paper series,'' \emph{Central bank
  liquidity Provision, risk-taking and economic efficiency}, no. 1542, 2013.

\bibitem{GervaisDecentralized}
A.~Gervais, G.~Karame, S.~Capkun, and V.~Capkun, ``Is bitcoin a decentralized
  currency?'' in \emph{IEEE Security and Privacy}, 2014.

\bibitem{Karame:2015:MBS:2786062.2732196}
\BIBentryALTinterwordspacing
G.~O. Karame, E.~Androulaki, M.~Roeschlin, A.~Gervais, and S.~\v{C}apkun,
  ``Misbehavior in bitcoin: A study of double-spending and accountability,''
  \emph{ACM Trans. Inf. Syst. Secur.}, vol.~18, no.~1, pp. 2:1--2:32, May 2015.
  [Online]. Available: \url{http://doi.acm.org/10.1145/2732196}
\BIBentrySTDinterwordspacing

\bibitem{cryptoeprint:2015:496}
A.~Gervais, H.~Ritzdorf, M.~Lucic, and S.~Capkun, ``Quantifying location
  privacy leakage from transaction prices,'' Cryptology ePrint Archive, Report
  2015/496, 2015, \url{http://eprint.iacr.org/}.

\bibitem{DBLP:journals/iacr/SompolinskyZ13}
Y.~Sompolinsky and A.~Zohar, ``Accelerating bitcoin's transaction processing.
  fast money grows on trees, not chains,'' \emph{{IACR} Cryptology ePrint
  Archive}, vol. 2013, p. 881, 2013.

\bibitem{DBLP:conf/fc/SompolinskyZ15}
\BIBentryALTinterwordspacing
------, ``Secure high-rate transaction processing in bitcoin,'' in
  \emph{Financial Cryptography and Data Security - 19th International
  Conference, {FC} 2015, San Juan, Puerto Rico, January 26-30, 2015, Revised
  Selected Papers}, 2015, pp. 507--527. [Online]. Available:
  \url{http://dx.doi.org/10.1007/978-3-662-47854-7_32}
\BIBentrySTDinterwordspacing

\bibitem{bitcointalk}
\BIBentryALTinterwordspacing
 [Online]. Available: \url{https://bitcointalk.org/index.php?topic=279249}
\BIBentrySTDinterwordspacing

\bibitem{gibson2016}
\BIBentryALTinterwordspacing
A.~Gibson, ``An investigation into confidential transactions.'' [Online].
  Available:
  \url{https://github.com/AdamISZ/ConfidentialTransactionsDoc/blob/master/essayonCT.pdf}
\BIBentrySTDinterwordspacing

\bibitem{Zdanowicz:2004:DML:986213.986239}
\BIBentryALTinterwordspacing
J.~S. Zdanowicz, ``Detecting money laundering and terrorist financing via data
  mining,'' \emph{Commun. ACM}, vol.~47, no.~5, pp. 53--55, May 2004. [Online].
  Available: \url{http://doi.acm.org/10.1145/986213.986239}
\BIBentrySTDinterwordspacing

\bibitem{mimblewimble}
T.~E. Jedusor, ``Mimblewimble,'' July 2016,
  \url{https://download.wpsoftware.net/bitcoin/wizardry/mimblewimble.txt}.

\bibitem{cve}
\BIBentryALTinterwordspacing
 [Online]. Available: \url{https://cve.mitre.org/data/downloads/allitems.html}
\BIBentrySTDinterwordspacing

\bibitem{blackhat}
\BIBentryALTinterwordspacing
 [Online]. Available:
  \url{https://media.blackhat.com/bh-dc-11/Grand/BlackHat_DC_2011_Grand-Workshop.pdf}
\BIBentrySTDinterwordspacing

\bibitem{Hamdioui:2014:HPI:2616606.2616728}
\BIBentryALTinterwordspacing
S.~Hamdioui, G.~Di~Natale, G.~van Battum, J.-L. Danger, F.~Smailbegovic, and
  M.~Tehranipoor, ``Hacking and protecting ic hardware,'' in \emph{Proceedings
  of the Conference on Design, Automation \& Test in Europe}, ser. DATE
  '14.\hskip 1em plus 0.5em minus 0.4em\relax 3001 Leuven, Belgium, Belgium:
  European Design and Automation Association, 2014, pp. 99:1--99:7. [Online].
  Available: \url{http://dl.acm.org/citation.cfm?id=2616606.2616728}
\BIBentrySTDinterwordspacing

\bibitem{Rivest2001}
\BIBentryALTinterwordspacing
R.~L. Rivest, A.~Shamir, and Y.~Tauman, \emph{How to Leak a Secret}.\hskip 1em
  plus 0.5em minus 0.4em\relax Berlin, Heidelberg: Springer Berlin Heidelberg,
  2001, pp. 552--565. [Online]. Available:
  \url{http://dx.doi.org/10.1007/3-540-45682-1_32}
\BIBentrySTDinterwordspacing

\bibitem{DBLP:journals/corr/SompolinskyZ16}
\BIBentryALTinterwordspacing
Y.~Sompolinsky and A.~Zohar, ``Bitcoin's security model revisited,''
  \emph{CoRR}, vol. abs/1605.09193, 2016. [Online]. Available:
  \url{http://arxiv.org/abs/1605.09193}
\BIBentrySTDinterwordspacing

\bibitem{DBLP:conf/atal/LewenbergBSZR15}
\BIBentryALTinterwordspacing
Y.~Lewenberg, Y.~Bachrach, Y.~Sompolinsky, A.~Zohar, and J.~S. Rosenschein,
  ``Bitcoin mining pools: {A} cooperative game theoretic analysis,'' in
  \emph{Proceedings of the 2015 International Conference on Autonomous Agents
  and Multiagent Systems, {AAMAS} 2015, Istanbul, Turkey, May 4-8, 2015}, 2015,
  pp. 919--927. [Online]. Available:
  \url{http://dl.acm.org/citation.cfm?id=2773270}
\BIBentrySTDinterwordspacing

\bibitem{DBLP:journals/corr/Kokoris-KogiasJ16}
\BIBentryALTinterwordspacing
E.~Kokoris{-}Kogias, P.~Jovanovic, N.~Gailly, I.~Khoffi, L.~Gasser, and
  B.~Ford, ``Enhancing bitcoin security and performance with strong consistency
  via collective signing,'' \emph{CoRR}, vol. abs/1602.06997, 2016. [Online].
  Available: \url{http://arxiv.org/abs/1602.06997}
\BIBentrySTDinterwordspacing

\bibitem{jute}
\BIBentryALTinterwordspacing
 [Online]. Available:
  \url{https://blog.sia.tech/2016/05/14/towards-a-sub-second-block-size/}
\BIBentrySTDinterwordspacing

\bibitem{stellar}
\BIBentryALTinterwordspacing
 [Online]. Available: \url{https://www.stellar.org}
\BIBentrySTDinterwordspacing

\bibitem{tendermint}
\BIBentryALTinterwordspacing
 [Online]. Available: \url{http://tendermint.com}
\BIBentrySTDinterwordspacing

\bibitem{ycom}
\BIBentryALTinterwordspacing
 [Online]. Available: \url{https://news.ycombinator.com/item?id=9342348}
\BIBentrySTDinterwordspacing

\bibitem{lightning}
\BIBentryALTinterwordspacing
J.~Poon and T.~Dryja, ``The bitcoin lightning network: Scalable off-chain
  instant payments,'' 2015. [Online]. Available:
  \url{http://lightning.network/lightning-network-paper.pdf}
\BIBentrySTDinterwordspacing

\bibitem{194906}
\BIBentryALTinterwordspacing
I.~Eyal, A.~E. Gencer, E.~G. Sirer, and R.~V. Renesse, ``Bitcoin-ng: A scalable
  blockchain protocol,'' in \emph{13th USENIX Symposium on Networked Systems
  Design and Implementation (NSDI 16)}.\hskip 1em plus 0.5em minus 0.4em\relax
  Santa Clara, CA: USENIX Association, Mar. 2016, pp. 45--59. [Online].
  Available:
  \url{https://www.usenix.org/conference/nsdi16/technical-sessions/presentation/eyal}
\BIBentrySTDinterwordspacing

\bibitem{DBLP:conf/sss/DeckerW15}
\BIBentryALTinterwordspacing
C.~Decker and R.~Wattenhofer, ``A fast and scalable payment network with
  bitcoin duplex micropayment channels,'' in \emph{Stabilization, Safety, and
  Security of Distributed Systems - 17th International Symposium, {SSS} 2015,
  Edmonton, AB, Canada, August 18-21, 2015, Proceedings}, 2015, pp. 3--18.
  [Online]. Available: \url{http://dx.doi.org/10.1007/978-3-319-21741-3_1}
\BIBentrySTDinterwordspacing

\bibitem{Sirer:2016:TPS:2907055.2896382}
\BIBentryALTinterwordspacing
E.~G. Sirer, ``Technical perspective: The state (and security) of the bitcoin
  economy,'' \emph{Commun. ACM}, vol.~59, no.~4, pp. 85--85, Mar. 2016.
  [Online]. Available: \url{http://doi.acm.org/10.1145/2896382}
\BIBentrySTDinterwordspacing

\bibitem{Andrychowicz:2016:SMC:2907055.2896386}
\BIBentryALTinterwordspacing
M.~Andrychowicz, S.~Dziembowski, D.~Malinowski, and L.~Mazurek, ``Secure
  multiparty computations on bitcoin,'' \emph{Commun. ACM}, vol.~59, no.~4, pp.
  76--84, Mar. 2016. [Online]. Available:
  \url{http://doi.acm.org/10.1145/2896386}
\BIBentrySTDinterwordspacing

\end{thebibliography}
\end{document}